\documentstyle[aps,twocolumn,epsf]{revtex}
\begin{document}
\draft
\title
{Inequalities relating area, energy, surface gravity and charge of black holes}
\author{Sean A. Hayward}
\address{Yukawa Institute for Theoretical Physics, Kyoto University,
Kitashirakawa, Sakyo-ku, Kyoto 606-8502, Japan\\
\tt hayward@yukawa.kyoto-u.ac.jp}
\date{29th June 1998}
\maketitle

\begin{abstract}
The Penrose-Gibbons inequality for charged black holes is proved 
in spherical symmetry,
assuming that outside the black hole there are no current sources,
meaning that the charge $e$ is constant,
with the remaining fields satisfying the dominant energy condition.
Specifically, for any achronal hypersurface 
which is asymptotically flat at spatial or null infinity 
and has an outermost marginal surface of areal radius $r$,  
the asymptotic mass $m$ satisfies $2m\ge r+e^2/r$.
Replacing $m$ by a local energy $\mu$, 
the inequality holds locally outside the black hole.
A recent definition of dynamic surface gravity $\kappa$ 
also satisfies inequalities $2\kappa\le1/r-e^2/r^3$
and $m\ge\mu\ge r^2\kappa+e^2/r$.
All these inequalities are sharp in the sense that 
equality is attained for the Reissner-Nordstr\"om black hole.

\end{abstract}

\section{Introduction}
In general relativity, one expects that black holes will form 
if matter or energy is sufficiently concentrated,
and conversely that a black hole of given size will require 
a certain minimal amount of energy.
The latter expectation was formulated by Penrose\cite{P}
in the form of an inequality 
\begin{equation}
m\ge\sqrt{A/16\pi}
\end{equation}
where $A$ is the area of the black hole,
$m$ the mass or energy at spatial infinity, 
assuming an asymptotically flat space-time,
and the Newtonian gravitational constant is unity.
Penrose argued that this was required by his cosmic censorship hypothesis.
Although the Penrose inequality remains a conjecture in general,
it has recently been established for time-symmetric hypersurfaces\cite{HI}
and in spherical symmetry\cite{ge}.
The latter result also holds locally outside the black hole 
with $m$ replaced by a local energy $E$,
so that one need not assume asymptotic flatness.
This proof and localization of the Penrose inequality in spherical symmetry
was first reported as a special case 
of an inequality for the Hawking energy\cite{mono}
and seems to be widely unknown.

Moreover, 
there is another inequality relating the area and surface gravity $\kappa$
as recently defined for a dynamic, spherically symmetric black hole\cite{1st}:
\begin{equation}
\kappa\le\sqrt{\pi/A}.
\end{equation}
Physically this means that for a black hole of given size,
the gravitational attraction (acceleration) cannot exceed a certain amount.

The two inequalities combine to yield a third inequality
\begin{equation}
4\pi m\ge A\kappa
\end{equation}
which also holds in Newtonian gravity.
Physically this means that the total surface gravity 
(integrated over the surface) 
cannot exceed a value given by the asymptotic mass.
Again this may be localized in terms of $E$.

All these inequalities require the dominant energy condition.
They are also sharp in the sense that equality is attained 
for the Schwarzschild black hole.
However, 
equality is not attained for the Reissner-Nordstr\"om charged black hole.
Gibbons\cite{G} has therefore suggested that for charged black holes,
there should be a sharpened version of the Penrose inequality, 
relating the area, mass and charge.
This Gibbons inequality is established here in spherical symmetry,
along with sharpened versions of the other two inequalities
and localized versions with $m$ replaced (where it appears) 
by a local energy $\mu$.

\section{Geometry}
In spherical symmetry, 
the area $A$ of the spheres of symmetry is a geometrical invariant.
It is also convenient to use the areal radius $r=\sqrt{A/4\pi}$.
A sphere is said to be {\sl untrapped, marginal or trapped}
as $dA$ is spatial, null or temporal respectively,
where $d$ is the exterior derivative.
The covariant and contravariant duals induced by the space-time metric 
(index lowering and raising) will not be denoted explicitly.

A marginal sphere will be taken as the local characterization of a black hole.
The more precise conditions needed to distinguish 
black holes from white holes, inner horizons or wormholes\cite{ge,1st,bhd,wh} 
will not be relevant in this article.
The results will be stated in terms of the area of a marginal sphere,
but can be extended for trapped spheres inside the black hole,
since area decreases in future-causal directions in future-trapped regions.

Untrapped or marginal spheres have a spatial orientation 
determined by the fact that the area increases 
in any spatial direction to one side of the sphere,
and decreases to the other side\cite{ge,mono,1st,bhd}.
These will be called {\sl outward}\/ and {\sl inward}\/ directions respectively.
Specifically, $z$ will henceforth denote an outward achronal vector
on an untrapped or marginal sphere, 
where achronal will be used to mean spatial or null.
Then
\begin{equation}
A'>0
\end{equation}
where $f'=z\cdot df$, the dot denoting contraction.

Ref.\cite{1st} introduced two invariants of the energy tensor $T$,
the function
\begin{equation}
w=-\textstyle{1\over2}\hbox{trace}\,T
\end{equation}
and the vector
\begin{equation}
\psi=T\cdot dr+wdr
\end{equation}
where the trace refers to the two-dimensional space 
normal to the spheres of symmetry
and the sign convention is that spatial metrics are positive definite.
One may physically interpret $w$ as an energy density 
and $\psi$ as an energy flux,
actually a local version of the (outward minus inward) Bondi flux.
Assuming the dominant energy condition, 
\begin{equation}
w\ge0.
\end{equation}
Assuming only the null energy condition, 
$\psi$ is outward achronal (or zero) in untrapped regions, implying that 
\begin{equation}
z\cdot\psi\ge0.
\end{equation}
These two facts are the basis for the Penrose and Gibbons inequalities.

\section{Penrose inequality}
The Misner-Sharp energy\cite{MS} may be defined by
\begin{equation}
E=\textstyle{1\over2}r\left(1-dr\cdot dr\right).
\end{equation}
Then the gradient of $E$ is determined by the Einstein equation as
\begin{equation}
dE=A(\psi+wdr).
\end{equation}
This equation was called the unified first law in Ref.\cite{1st}, 
since it encodes first laws of both black-hole dynamics 
and relativistic thermodynamics.

Projecting the unified first law along $z$,
the dominant energy condition therefore implies
\begin{equation}
E'\ge0.
\end{equation}
This is the monotonicity property of $E$ established in Ref.\cite{ge}
and generalized for the Hawking energy in Ref.\cite{mono}:
in an untrapped region, $E$ is non-decreasing in any outward achronal direction.
This immediately gives proofs of both the positive-mass theorem
and the Penrose inequality.
In the latter case, 
one considers an untrapped achronal hypersurface
bounded at the inward end by a marginal surface with $r=r_0$
and therefore $E={1\over2}r_0$.
Then
\begin{equation}
E\ge\textstyle{1\over2}r_0
\end{equation}
on the hypersurface.
The inequality therefore holds anywhere in the untrapped region 
achronally outwards from the marginal surface.
In particular, 
if the hypersurface is asymptotically flat at spatial or null infinity,
with asymptotic (ADM or Bondi) energy $m$,
then since $m$ is the asymptotic limit of $E$\cite{ge}, it also follows that
\begin{equation}
m\ge\textstyle{1\over2}r_0.
\end{equation}
This therefore established the Penrose inequality in spherical symmetry.
The result for $E$ is more general, applying locally outside the black hole 
with no need to assume asymptotic flatness.

\section{Gibbons inequality}
Turning to the Maxwell electromagnetic field:
a spherically symmetric
electromagnetic field tensor has only one independent component,
which can be encoded in either the charge $e$ or the electric field $e/r^2$,
the magnetic field vanishing\cite{1st}.
The Maxwell equations reduce to
\begin{equation}
de=AJ
\end{equation}
where the covector $J$ is the current density.
In the absence of a source $J$, $e$ is therefore constant.
The electromagnetic energy tensor $T_e$ has one independent component,
yielding
\begin{equation}
w_e=e^2/8\pi r^4\qquad\psi_e=0
\end{equation}
where the subscript $e$ refers to the electromagnetic field.
Therefore the unified first law can be written as
\begin{equation}
dE=A(\psi_o+w_odr)-\textstyle{1\over2}e^2d(r^{-1})
\end{equation}
where the subscript $o$ refers to other fields,
i.e.\ $T=T_e+T_o$ and so on.
Assuming that the other fields satisfy the dominant energy condition,
it follows that
\begin{equation}
2E'\ge-e^2(r^{-1})'.
\end{equation}
On an untrapped achronal hypersurface which contains no sources $J$,
is asymptotically flat (at spatial or null infinity) with asymptotic energy $m$ 
and is bounded by a marginal surface of areal radius $r_0$,
the inequality integrates to
\begin{equation}
2m\ge r_0+e^2/r_0
\end{equation}
which is the Gibbons inequality.
This therefore establishes the Gibbons inequality in spherical symmetry.
It should be stressed that this need not hold 
if there are sources $J$ outside the black hole, 
as already pointed out by Malec \& \'O Murchadha\cite{MM},
who proved the inequality for maximal hypersurfaces.
Iriondo et al.\cite{IMM} similarly proved the Penrose inequality
for constant-mean-curvature hypersurfaces.

The Gibbons inequality may be localized by introducing a local energy
\begin{equation}
\mu=E+e^2/2r
\end{equation}
which may be interpreted as the combined energy $E$ 
minus the electromagnetic contribution.
Then the unified first law can be rewritten
\begin{equation}
d\mu=A(\psi_o+w_odr)+\textstyle{1\over2}r^{-1}d(e^2)
\end{equation}
and implies
\begin{equation}
2\mu'\ge r^{-1}(e^2)'.
\end{equation}
The right-hand side vanishes in the absence of sources $J$,
in which case, integrating over an untrapped achronal hypersurface 
bounded at the inward end by a marginal surface of areal radius $r_0$,
\begin{equation}
2\mu\ge r_0+e^2/r_0.
\end{equation}
This is a local version of the Gibbons inequality,
holding anywhere in the untrapped region 
achronally outwards from the marginal surface,
with no need to assume asymptotic flatness.
Again it should be stressed that this has assumed 
no sources $J$ outside the black hole, 
though this could clearly be weakened 
to allow monotonically strengthening charge: $(e^2)'\ge0$.

\section{Inequalities involving surface gravity}
A definition of surface gravity $\kappa$ 
for dynamic, spherically symmetric black holes was proposed in Ref.\cite{1st} 
by substituting the Kodama vector for the Killing vector 
in the standard definition of surface gravity for stationary black holes.
This gives 
\begin{equation}
2\kappa={*}d{*}dr
\end{equation}
where $*$ is the Hodge operator of the two-dimensional space 
normal to the spheres of symmetry, i.e.\ ${*}d{*}$ is a divergence.
The Einstein equation implies
\begin{equation}
\kappa=E/r^2-4\pi rw.
\end{equation}
The dominant energy condition therefore implies
\begin{equation}
E\ge r^2\kappa.
\end{equation}
In Newtonian gravity this would be an equality,
expressing Newton's law of gravitation,
combined with the fact that a spherically symmetric body 
has the same external gravitational field as a point source with the same mass.
The inequality holds anywhere in the space-time 
and reduces on a marginal surface to
\begin{equation}
\kappa_0\le1/2r_0.
\end{equation}
Combining with the Penrose inequality yields
\begin{equation}
m\ge r_0^2\kappa_0.
\end{equation}
These three inequalities were established in Ref.\cite{1st}
and may be physically interpreted as in the Introduction.

In each case, equality is attained by the Schwarzschild black hole
but not by the Reissner-Nordstr\"om black hole,
but the inequalities may be sharpened for the electromagnetic case by expanding
\begin{equation}
\kappa=(\mu-e^2/r)/r^2-4\pi rw_o.
\end{equation}
Assuming that the other fields satisfy the dominant energy condition,
\begin{equation}
\mu\ge r^2\kappa+e^2/r
\end{equation}
which reduces on a marginal surface to
\begin{equation}
2\kappa_0\le1/r_0-e^2/r_0^3.
\end{equation}
This in turn may be combined with the Gibbons inequality to yield
\begin{equation}
m\ge r_0^2\kappa_0+e^2/r_0.
\end{equation}
All three inequalities attain equality for the Reissner-Nordstr\"om black hole.
The last two inequalities are the desired sharpened versions 
of the inequalities displayed in the Introduction,
while the first is a localized version of the last.

\section{Remarks}
All the inequalities derived here 
may be conjectured to hold even without spherical symmetry,
though there are currently no agreed general definitions of 
quasi-local energy or surface gravity. 

Research supported by a European Union Science and Technology Fellowship.

\end{document}